\documentstyle[sprocl]{article}

\def\be{\begin{equation}}
\def\ee{\end{equation}}
\def\bea{\begin{eqnarray}}
\def\eea{\end{eqnarray}}

\begin{document}
\title{WAVELET ANALYSES OF HIGH MULTIPLICITY EVENTS}
\author{M. BIYAJIMA, N. SUZUKI ${}^*$ and A. OHSAWA ${}^{**}$}
\address{Department of Physics, Faculty of Science, \\
Shinshu University, Matsumoto 390, Japan,\\
${}^{*}$ Matsusho Gakuen Junior College, Matsumoto 390-12, Japan, \\
${}^{**}$ Institute for Cosmic Ray Research, \\
University of Tokyo, Tanashi 188, Japan}
%
%
\maketitle\abstracts{
Pseudo-rapidity distributions of high multiplicity JACEE events 
( Ca-C, Si-AgBr and G27) and Texas Lone Star (TLS) event are 
analysed by the wavelet transform method. 
Using the Daubechies' waveltes, $D_4, D_6, D_8,$ and $ D_{10} $,   
wavelet spectra of those events are 
calculated. Two JACEE events are compared with the 
simulation calculations.  We  discuss how to distinguish  
prominent peak from apparent peaks. 
The wavelet spectrum of Ca-C event seems to resemble  that 
simulated with the Poisson random number distributions. 
%
%
The wavelet spectrum of TLS event shows an oscillatory behavior. 
It is found that the wavelet spectra of JACEE G27 event are different 
from those of Arizona group.}

\section{Introduction}
 In high energy nucleus-nucleus (AA) collisions number density of 
secondary particles in the rapidity space becomes very high and 
studies of number density fluctuations in the rapidity space is 
expected to reveal new features of multiparticle production 
mechanisms.\cite{taka84,suzu91,biya91,ptp95}
The pseudo-rapidity distributions of the three 
JACEE events \cite{burn83}, (Ca-C, Si-AgBr  and G27 ) and  the 
Texas Lone Star (TLS)~\cite{prs64} event
are analysed by the wavelet 
transform method~\cite{daub88,stra89,newl93,mori88}. \\
  Any function (data) can be expanded into self-similar wavepackets 
in this scheme.  
Wavelet spectra of those four events 
are calculated and some of them are  compared with 
simulation calculations.\cite{ptp95}  In 
$\S$ 2 the wavelet transform concept is introduced  
and in $\S$ 3 the Daubechies' mother waveltes, $D_4, D_6, D_8,$ 
and $ D_{10} $, 
are prepared to be used in $\S$ 4. The wavelet spectra of the 
above mentioned four events are calculated and compared with the 
simulation calculations in $\S$ 4.
The final section is devoted to the concluding 
remarks.
%
%
%
\section{Wavelet transform}
 Wavelets are constructed from dilation and translation of a 
scaling function $\phi(x)$, which is constructed 
by an iteration equation,
 \begin{equation}
         \phi_i(x)= \sum_{k=0}^{N-1} c_{k}\phi_{i-1}(2x-k)    
            \quad i=1,2,\cdots,         \label{eq:wvlt1}
 \end{equation}
\clearpage 
\noindent 
from a primary scaling function $\phi_0(x)$ 
( $\phi_{0}(x)=1$ for $0 \le x < 1$ and $\phi_{0}(x)=0$ otherwise).
In Eq.(\ref{eq:wvlt1}), N is a even number and $c_k$ 
$\,(k=0,1,\cdots,N-1)$ are constants.   The 
iteration is continued until $\phi_{i}(x)$ becomes 
indistinguishable 
from $\phi_{i-1}(x)$ and defines then $\phi(x)$. 

The mother wavelet $W(x)$ is given by 
  \begin{eqnarray}
      W(x) = \sum_{k=0}^{N-1} (-1)^{k+1} c_{N-1-k}\, \phi(2x+k-N+1),  
\label{eq:wvlt2}
  \end{eqnarray}
and defines the j-th level wavelet $(j=0,1,\cdots)$
  \begin{equation}
       \psi_{j,k}(x) = 2^{\frac{j}{2}} W(2^{j}x-k),   \qquad
                       k=0,1,\cdots,2^{j}-1.  \label{eq:wvlt3}
  \end{equation}
Coefficients $c_k\,(k=0,1,\cdots,N-1)$ should be determined in such a way  
that wavelets and scaling function satisfy the following 
conditions,
  \begin{eqnarray}
      \int \phi(x)\,\phi(x) dx &=& 1,    \quad 
  \int \phi(x)\,\psi_{j,k}(x) dx = 0,  \nonumber \\
  \int \psi_{j,k}(x)\,\psi_{r,s}(x) dx &=& \delta_{jr}\delta_{ks},\quad
   \int x^k W(x) dx = 0, \, k=0,\cdots,N/2-1.
                            \label{eq:wvlt4} 
  \end{eqnarray}
If N=4, we have
 \begin{eqnarray*}
      c_0 = \frac{1+\sqrt{3}}{4}, \quad
      c_1 = \frac{3+\sqrt{3}}{4}, \quad   
      c_2 = \frac{3-\sqrt{3}}{4}, \quad
      c_3 = \frac{1-\sqrt{3}}{4}.           
 \end{eqnarray*}
Then arbitrary function $f(x)$ defined in the region $0 \le x < 1$ 
can be  expanded as
 \begin{eqnarray*}
       f = a_{0}\phi(x) + \sum_{j=0}^{\infty}\,
            \sum_{k=0}^{2^{j}-1} \alpha_{j,k} \psi_{j,k}(x).
 \end{eqnarray*}

We shall consider here the case in which function $f(x)$ 
 $\,(0\le x <1)$ is given by its discrete values, 
$ f_i = f(\Delta/2 + \Delta \cdot (i-1))$, $\, (i=1,2,\cdots,2^r)$, 
where $\Delta \cdot 2^r = 1$.
By the use of column matrices,
 \begin{eqnarray*}
         F = \left( \begin{array}{c}
                   f_{1}         \\
                   f_{2}         \\
                   \cdots        \\
                   f_{2^r}
                     \end{array} \right),  \qquad\quad
        A_{j} = \left( \begin{array}{c}
                        \alpha_{j,0}  \\
                        \alpha_{j,1}  \\
                        \cdots        \\
                        \alpha_{j,2^{j}-1}
                       \end{array} \right),    \qquad j=0,1,\cdots,
 \end{eqnarray*}
the wavelet coefficients can be written as  
  \begin{eqnarray}
      a_{0} &=& 2^{-\frac{r}{2}}L_{1}L_{2}\cdots L_{r}F, 
\nonumber   \\
    A_{j-1} &=& 2^{-\frac{r}{2}}H_{j}L_{j+1}\cdots L_{r}F, 
           \quad j=1,2,\cdots,r-1,        \nonumber \\
        A_{r-1} &=& 2^{-\frac{r}{2}}H_{r}F.     \label{eq:wvlt5}       
  \end{eqnarray}
where $L_j$ and $H_j$ are $2^{j-1}\times 2^{j}$ matrices. 
The ${\it i}-$th row of $L_j$ is expressed as,
 \begin{eqnarray*}
      \left( \begin{array}{cccccccccc}
    0 & \cdots & 0 & l_{2(i-1)+1} & l_{2(i-1)+2} & \cdots &
          l_{2(i-1)+N} & 0 &  \cdots & 0
          \end{array} \right)          \nonumber  \\
      =  \frac{1}{\sqrt{2}}
      \left( \begin{array}{cccccccccc}
       0 & \cdots & 0 & c_{0} & c_{1} & \cdots &
          c_{N-1} & 0 &  \cdots & 0
          \end{array} \right).   
 \end{eqnarray*}
If ~$2(i-1)+k > 2^j $, element $l_{2(i-1)+k}$ is added to the 
$(2(i-1)+k-2^j)$ -th element in each row.
Matrix $H_j$ is obtained, if $c_k$ $\,(k=0,1,\cdots,N-1)$ in
 $L_j$ is replaced by $(-1)^{k+1}C_{N-1-k}$.

Matrices $L_j$ and $H_j$ satisfy the following conditions (coming from Eq. (\ref{eq:wvlt4})),
 \begin{eqnarray}
      L_j\,^{t}L_j = H_j\,^{t}H_j = I, \quad
      L_j\,^{t}H_j =  H_j\,^{t}L_j = 0,    
                               \label{eq:wvlt6}   
 \end{eqnarray}
where ${}^tL_j$ denotes the transpose of matrix $L_j$, and 
$I$ is the $2^{j-1}-$th order unit matrix.

From Eqs. (\ref{eq:wvlt5}) and (\ref{eq:wvlt6}), one gets the 
inverse wavelet transform: 
 \begin{eqnarray}
    F &=& F^\phi + \sum_{j=0}^{r-1} F^{(j)},   \nonumber  \\
   {}^{t}F^\phi &=& 2^{\frac{r}{2}}\,a_{0}L_{1}L_{2}\cdots L_{r},
         \nonumber \\
    {}^{t}F^{(j-1)} &=& 2^{\frac{r}{2}}\,\,
              {}^{t}A_{j-1}H_{j}L_{j+1}\cdots L_{r},  \qquad
               j=1,2,\cdots,r-1,  \nonumber  \\
    {}^{t}F^{(r-1)} &=& 2^{\frac{r}{2}}\,\,{}^{t}A_{r-1} H_{r},
         \label{eq:wvlt7} 
 \end{eqnarray}
Finally, Eqs. (\ref{eq:wvlt6}) and (\ref{eq:wvlt7}) lead to the following identity;
 \begin{eqnarray}
     {}^tF\,F\,2^{-r} &=& 2^{-r}\,\sum_{j=1}^{2^r} f_j^2
  = a_{0}^{2} + \sum_{j=0}^{r-1}E_{j}, 
          \nonumber \\
      E_{j} &=& {}^tA\,A = \sum_{k=0}^{2^{j}-1} \alpha_{j,k}^{2},
        \label{eq:wvlt8}
 \end{eqnarray}
where $E_j$ denotes the j-th level wavelet
spectrum \cite{yama91}.
\section{Daubechies' mother wavelets}
We prepare now Daubechies' mother wavelets, $D_4, D_6, D_8,$ and $ D_{10} $,
cf. Fig. 1, to be used for wavelet analysis below. 
If results are independent on these  mother wavelets, 
one can say that wavelet spectra bear some physical /or stochastic 
meaning to be further explored. 

\vspace*{7.0cm}
\begin{center}
{\small Fig. 1. Daubechies' mother wavelets, $D_4, D_6, D_8,$ 
and $ D_{10} $.}\\ 
\end{center}
\section{Wavelet spectra of the data}
 We analyse now by the wavelet 
transform \cite{daub88,stra89,newl93,mori88} the pseudo-rapidity 
distributions of the three 
JACEE events  and  the TLS event.

1)  In the case of Ca-C and Si-AgBr JACEE events (cf. Fig. 2), 
both original data and  subtracted data 
( original data - its background (BG)\cite{taka84}) are used. 
The corresponding wavelet spectra are shown in Fig. 2.
\clearpage
\vspace*{9.0cm}

\begin{center}
{\small Fig. 2. Ca-C and Si-AgBr JACEE events and their wavelet spectra.
Results obtained 
by the Poisson random number distributions are shown  
with $\langle E_j \rangle  + \sigma_j$ 
and $\langle E_j \rangle  - \sigma_j$. 
Notice that "$\langle ...\rangle$" 
means an assemble average.}\\
\end{center}

\noindent 
\\

We shall propose now how to distinguish a prominent peak from  
apparent ones. To this end, we use the wavelet coefficients satisfying the 
 following condition,
\begin{equation}
     | \alpha_{jk} - \mu | > \alpha \sigma,
\end{equation}
where $\mu$ is the average and $\sigma$ is the standard deviation 
of the wavelet coefficients.  
If $\alpha$ is chosen as 1.96/or 2.58,  the 
absolute values of the coefficients become large.  Our results show 
that  peak present in Si-AgBr event does not depend on the 
Daubechies' mother wavelets  $D_4, D_6, D_8,$ and $ D_{10} $. Therefore we 
can examine further its physical meaning. On the other hand, the peak in Ca-C 
event depends on the  Daubechies' mother wavelets.  Therefore we can say that
it is an apparent one. 
\clearpage

\vspace*{8.5cm}
\begin{center}
 {\small Fig. 3. Pseudo-rapidity distribution constructed by larger 
coefficients.}
\end{center} 

2) The TLS event is a photon distribution.  Using the 
same procedure,  we obtain its wavelet spectra shown in Fig. 4.  At present 
it is difficult to elucidate physical meaning of their oscillatory behaviour. 
\vspace*{5.5cm} 
\begin{center}
{\small Fig. 4.  Data  of TLS event and its wavelet spectra.} \\
\end{center}
\clearpage 
3) In this case of  JACEE event G27,  Arizona group \cite{ina96} and 
Bjorken \cite{bjo96} have conjectured  that this event probably contains 
some signal of the disoriented chiral condensate (DCC).
The ratio  $ r = n_0/( n_0 + n_{ch})$ \cite{huang96} is shown in Fig. 5.  
Using the same procedure as mentioned, we obtain the wavelet spectra shown 
in Fig. 5.
\vspace*{6.0cm}
\begin{center}
{\small Fig. 5.  JACEE G27 event and its wavelet spectra.}\\ 
\end{center}
It is found that our result is different from  that 
of Arizona group.\cite{ina96} 
(Notice that our method 
produces for the random distribution an increasing behaviour of wavelet 
spectrum when j increases.)  
\section {Concluding remarks}
We have prepared the Daubechies' mother wavelets, $D_4, D_6, 
D_8,$ and $ D_{10} $, and using them we have analysed 
high multiplicity events. 
It is found that our wavelet spectra obtained from three JACEE events 
and TLS event show different behaviour from that of 
Poisson random number distribution \cite{ptp95}.  
We conclude therefore that one can use the wavelet analysis to gain some
 physical information/or stochastic properties from 
the high multiplicity distributions.

\section*{Acknowledgments}
One of authors (M. B.) wishes to thank the Yamada Science 
Foundation for financial support for traveling expenses.
This work is partially 
supported by Japanese Grant-in-Aid for Scientific Research 
from the Ministry of Education, Science and Culture 
(No. 06640383) and (No. 08304024). 
N.S. thanks Matsusho Gakuen Junior College for the 
financial support. 

%

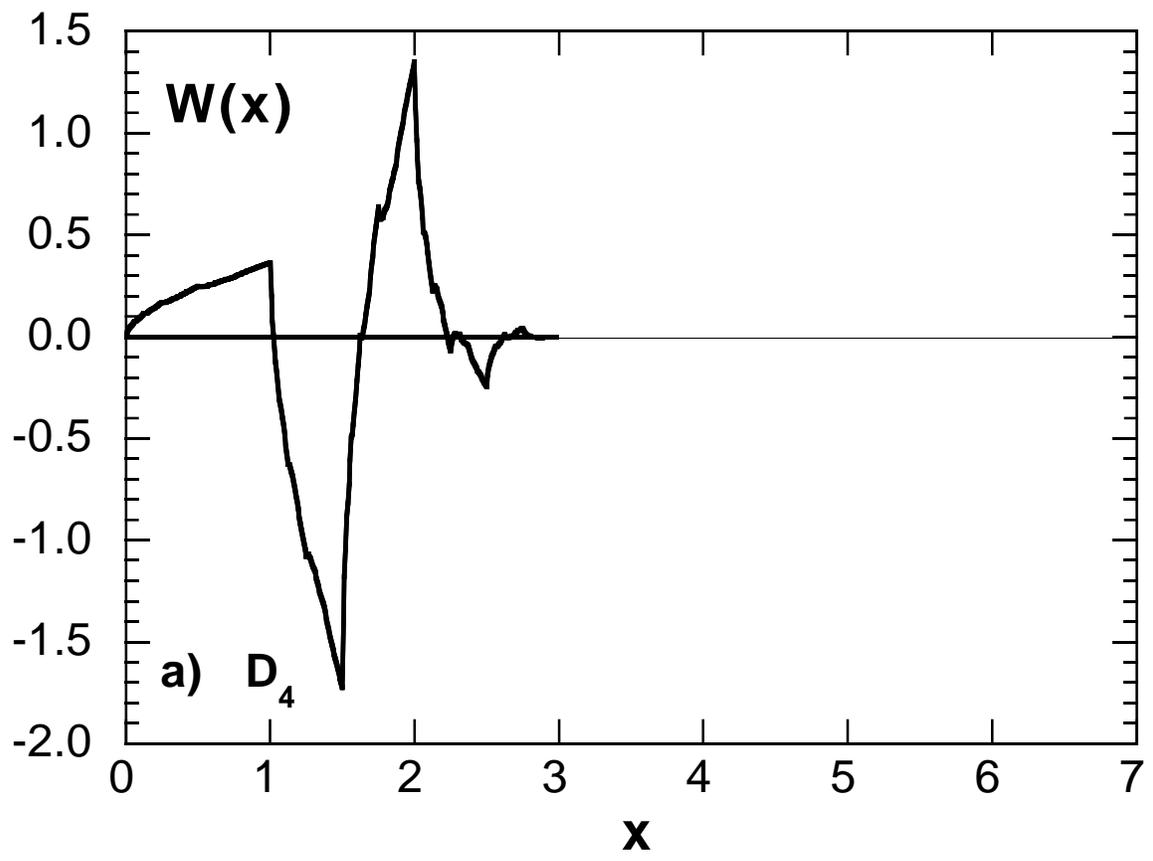
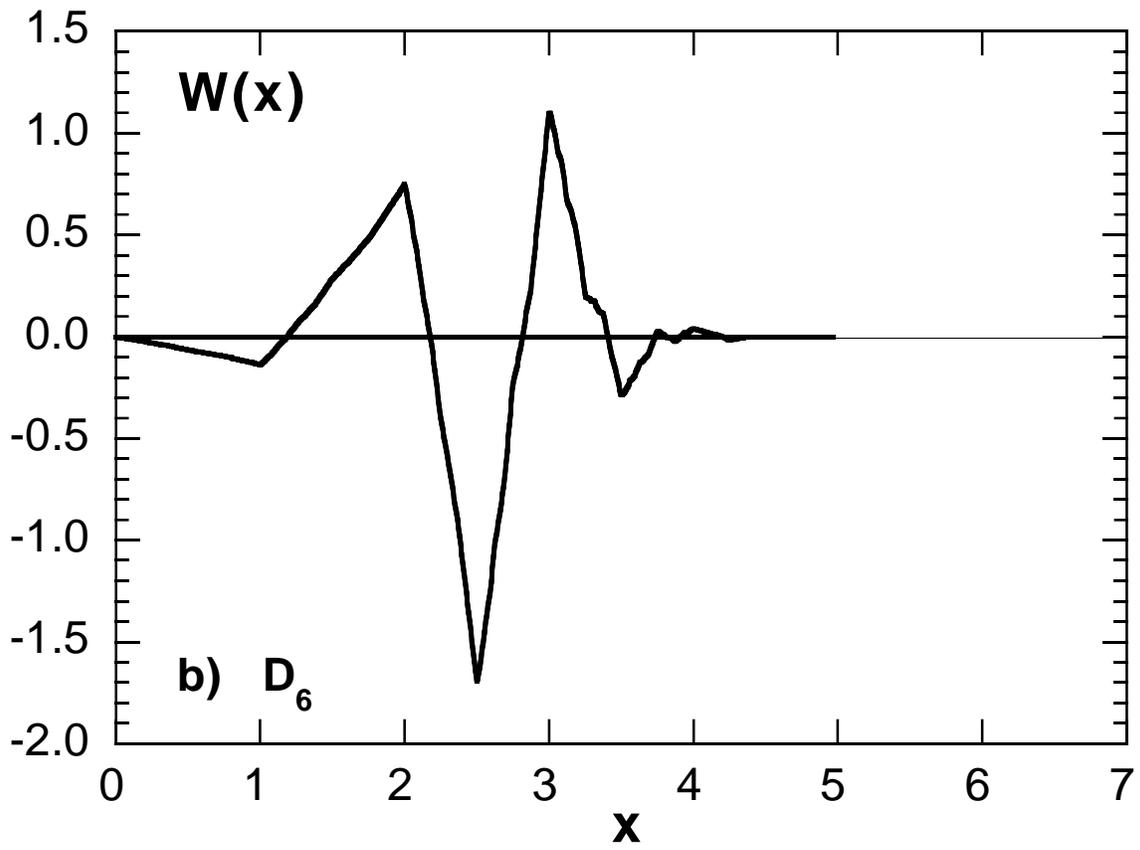
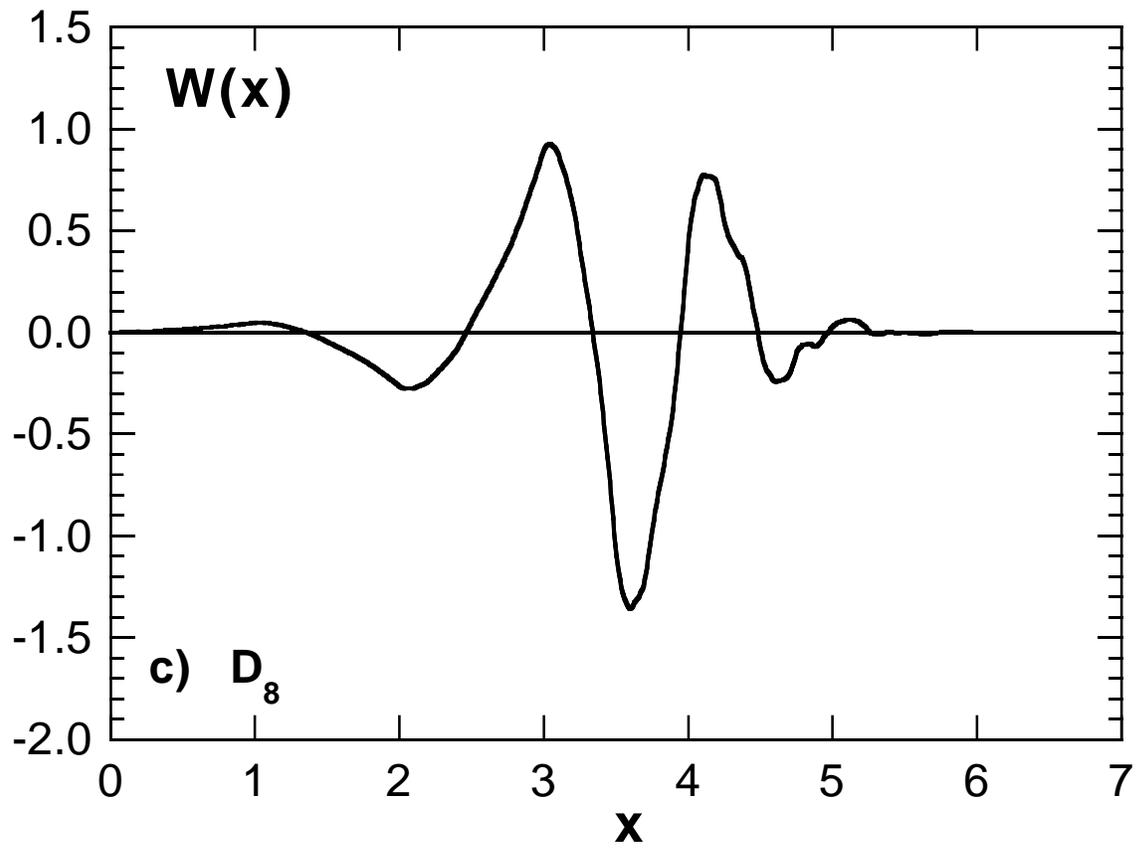
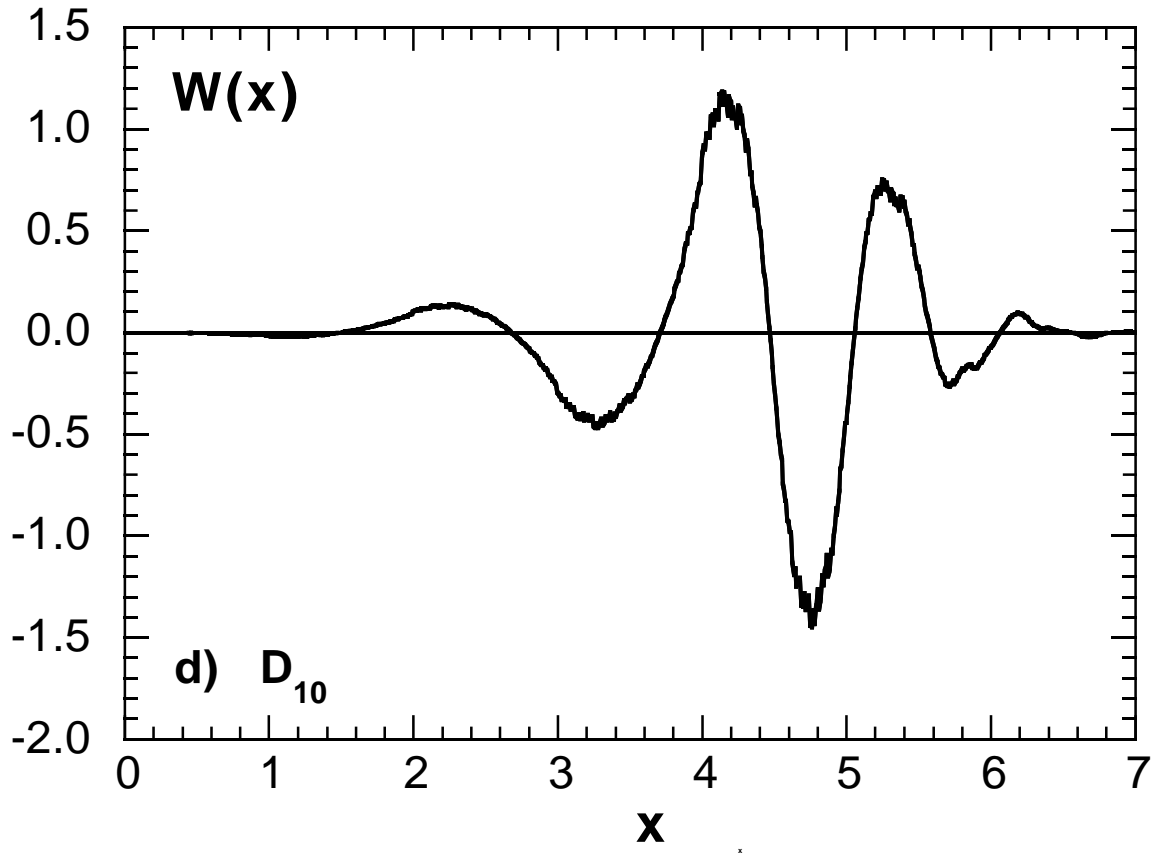
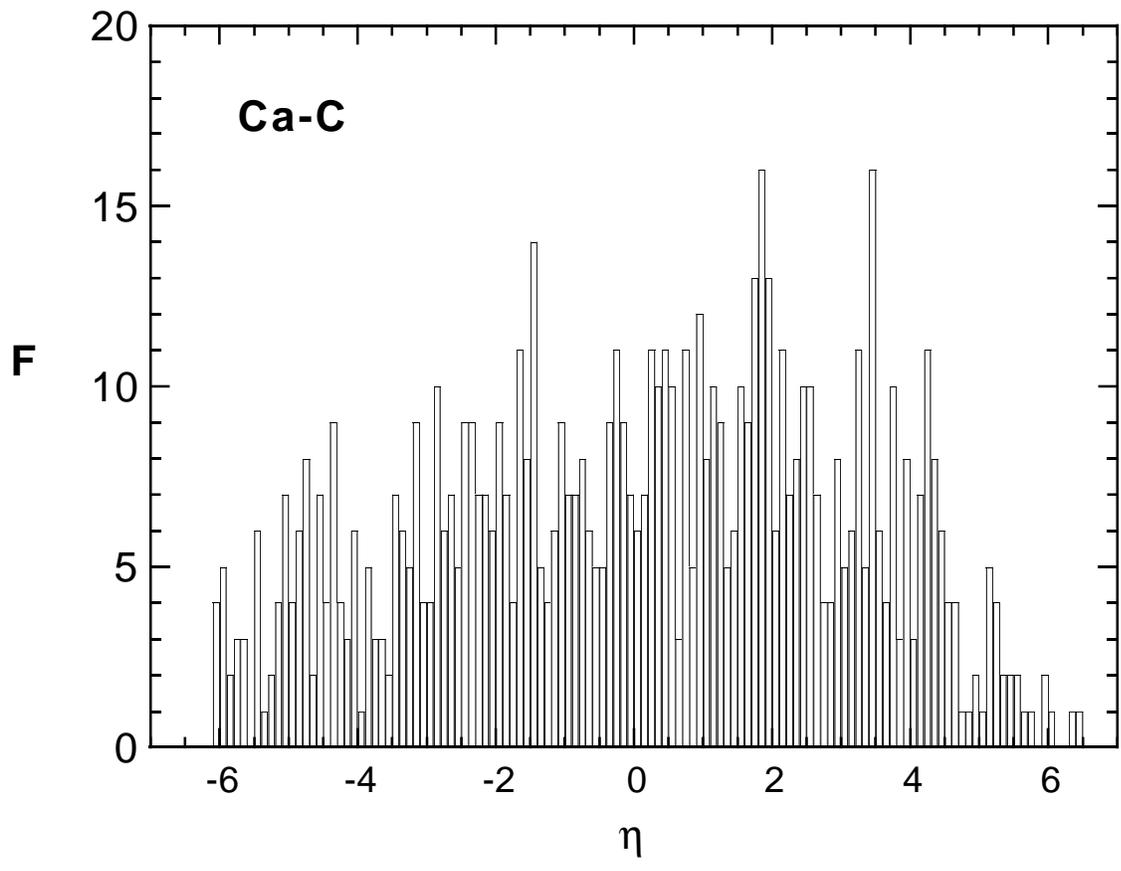
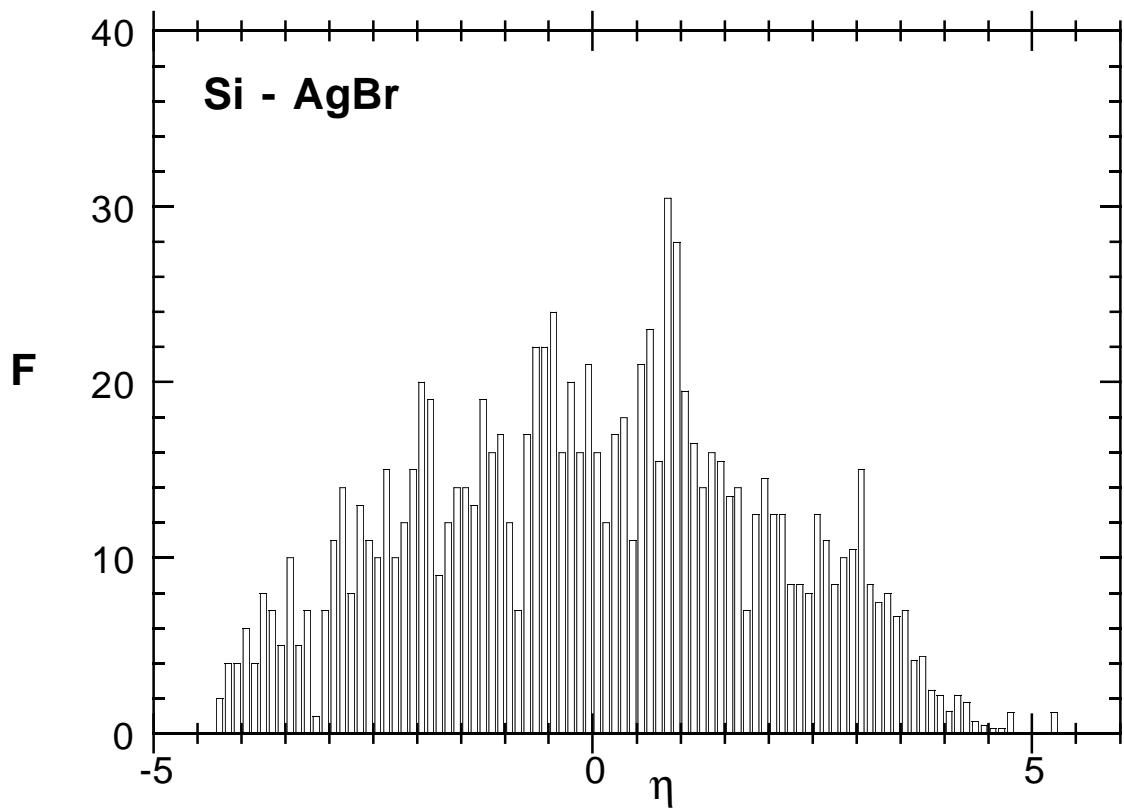
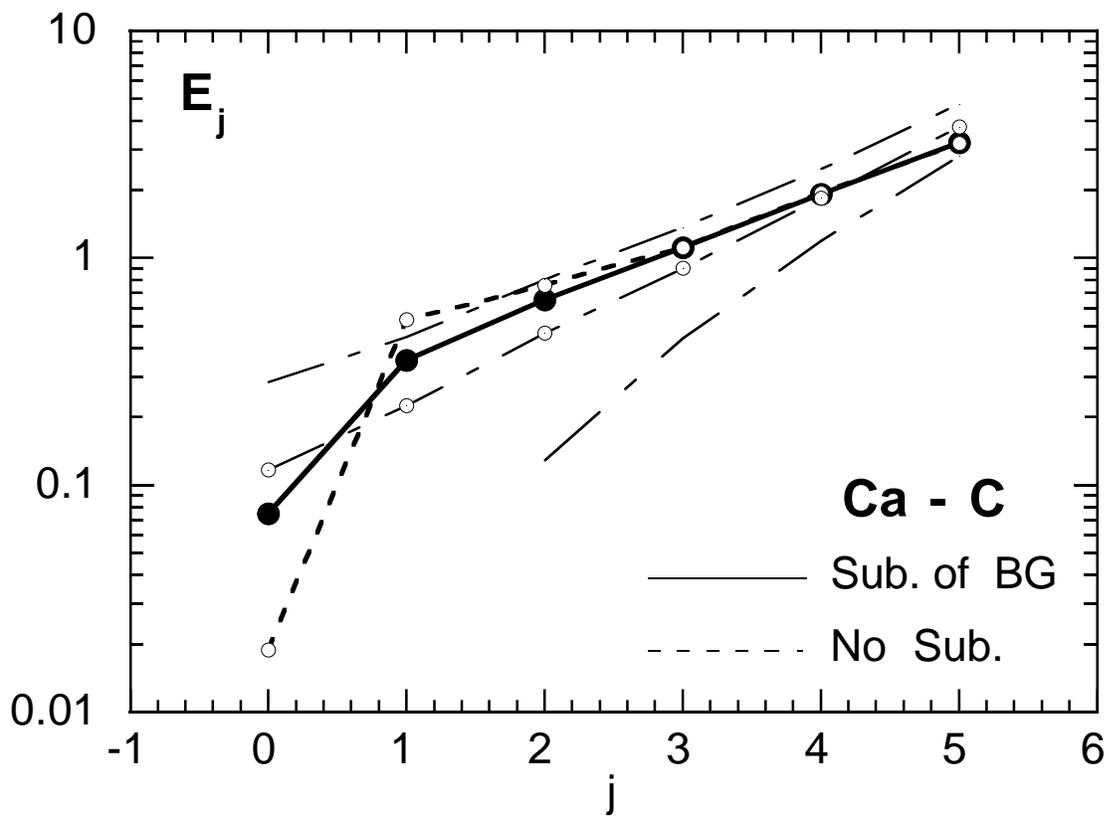
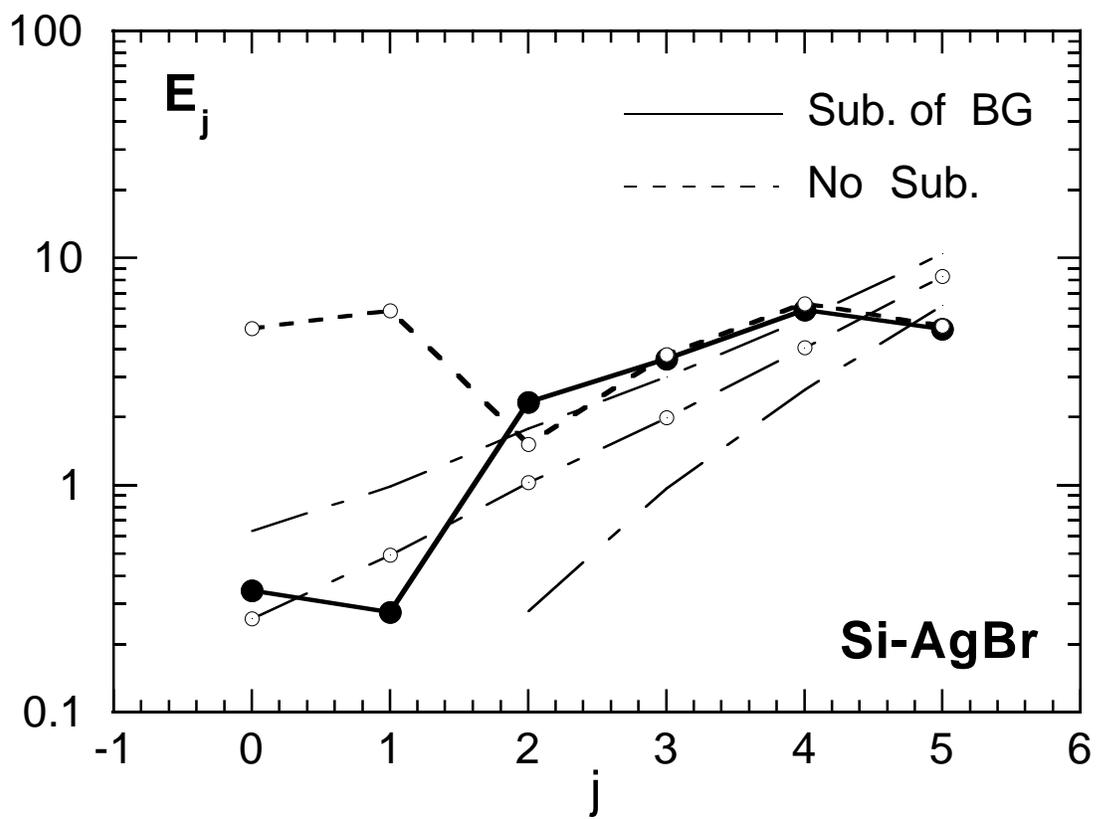
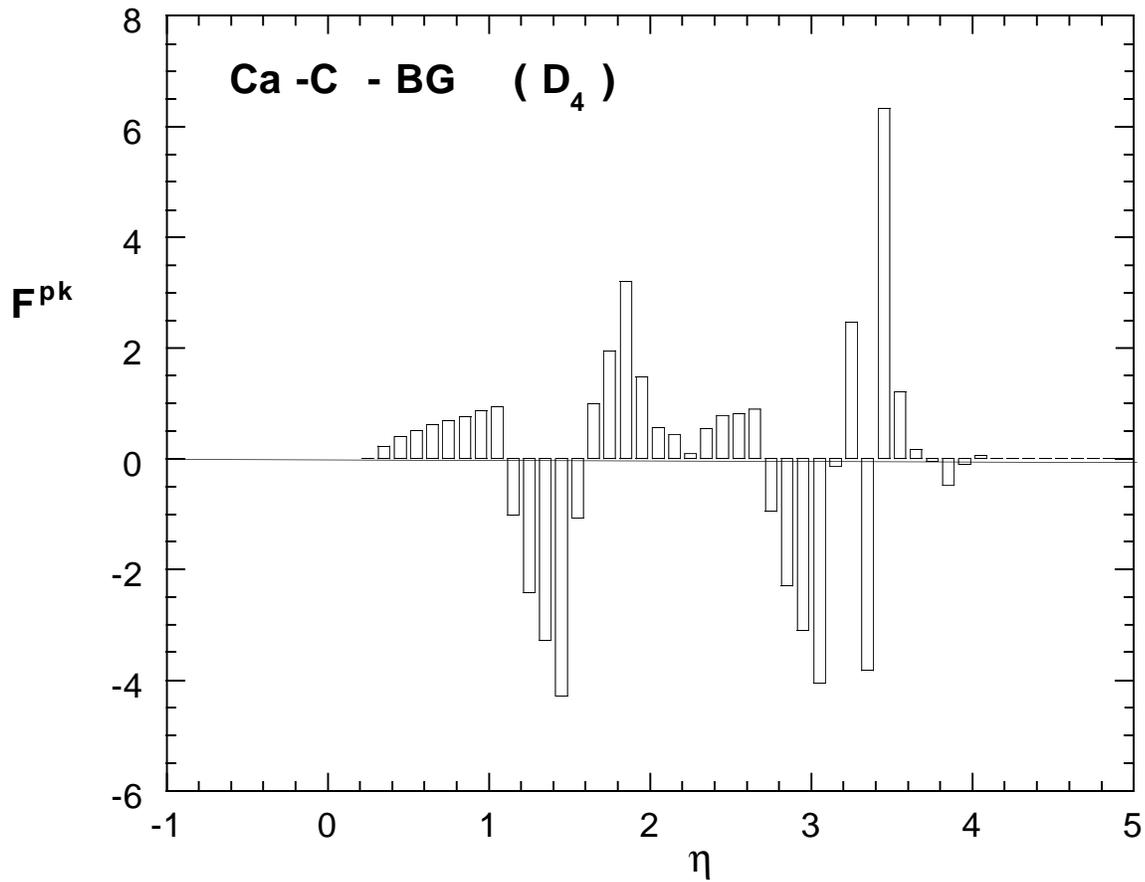
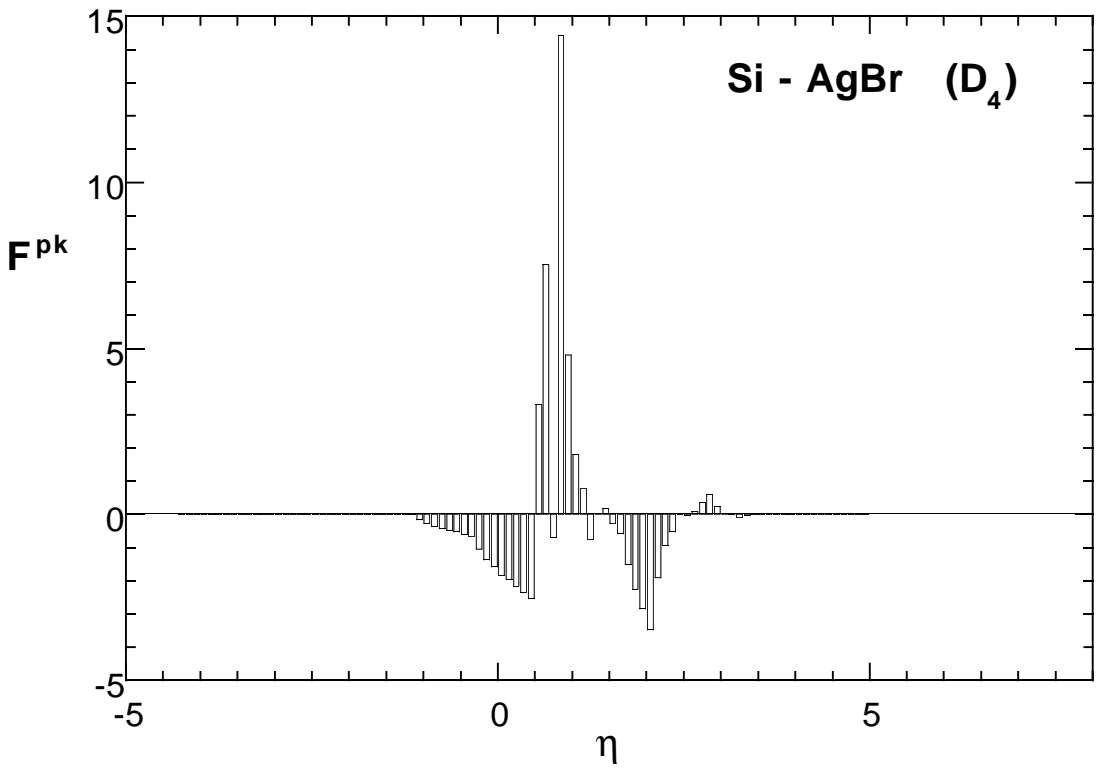
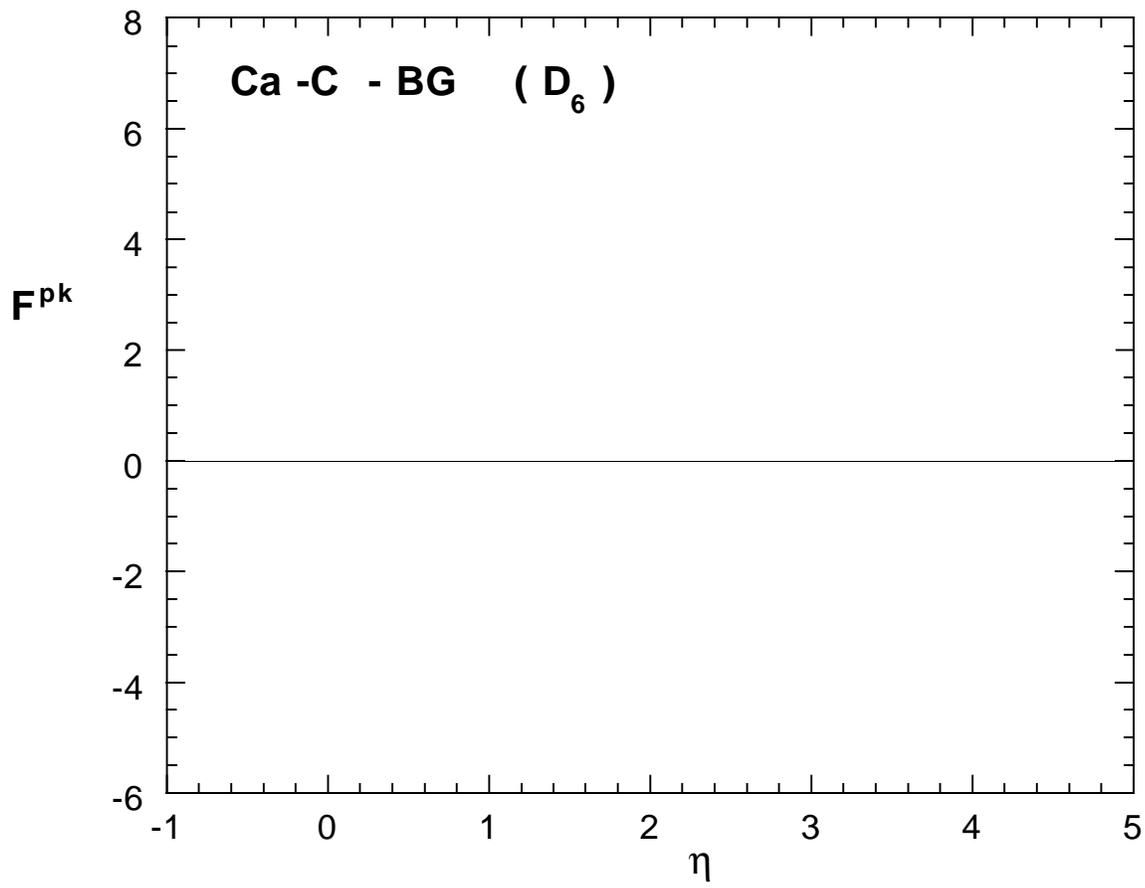
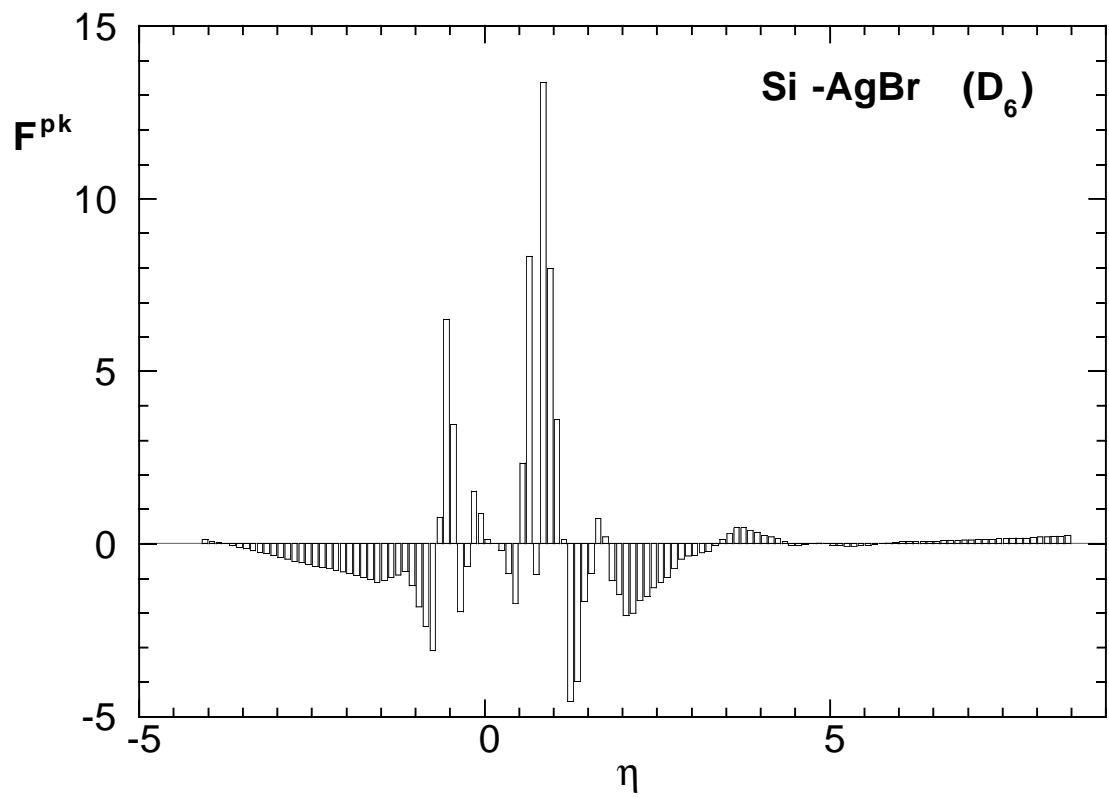
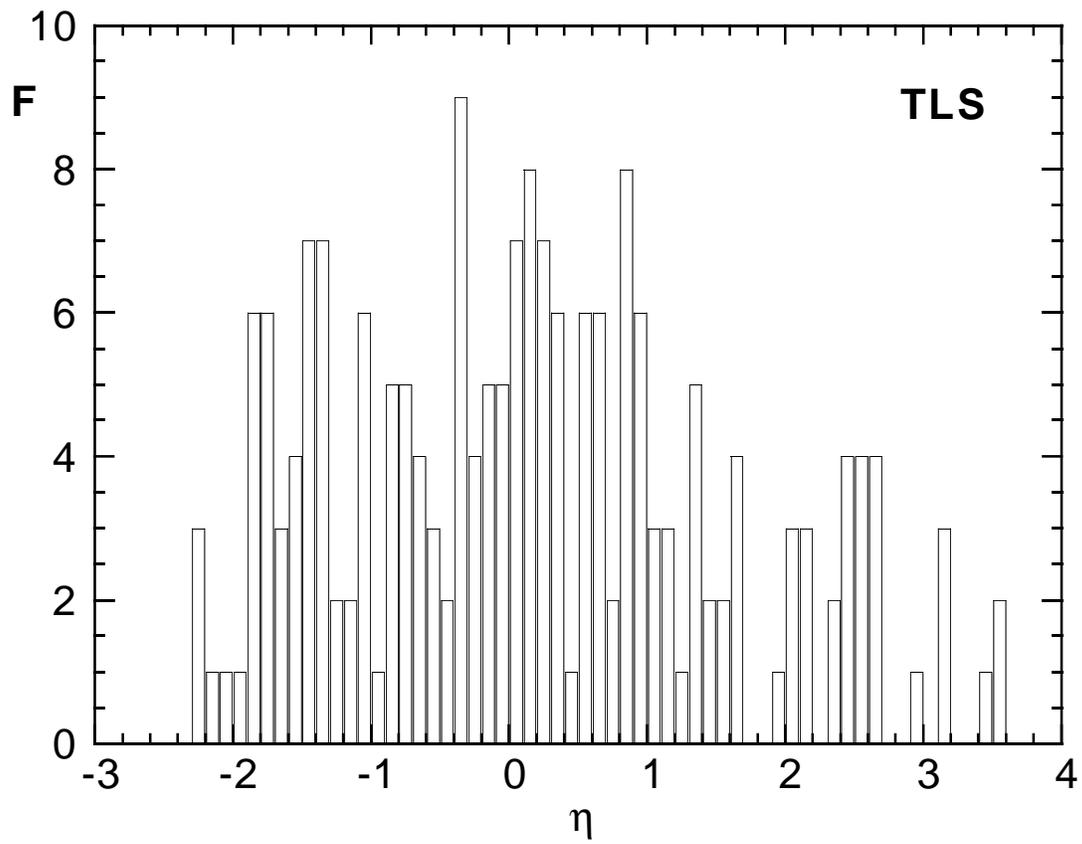
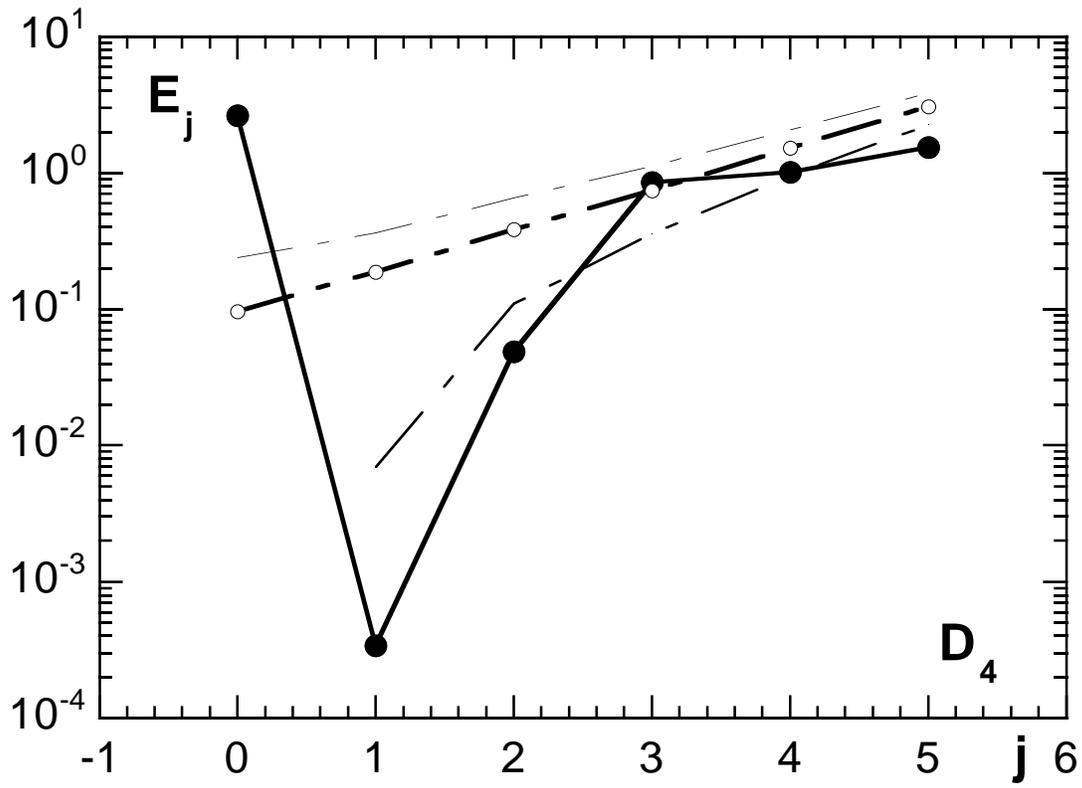
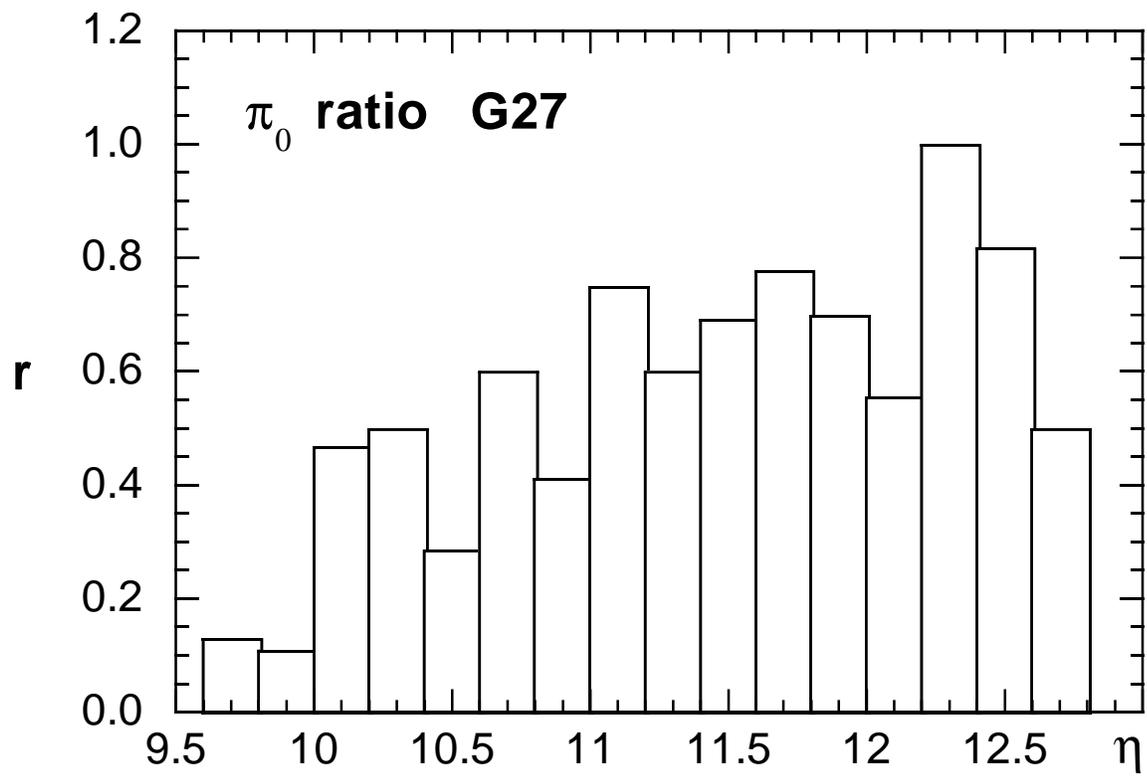
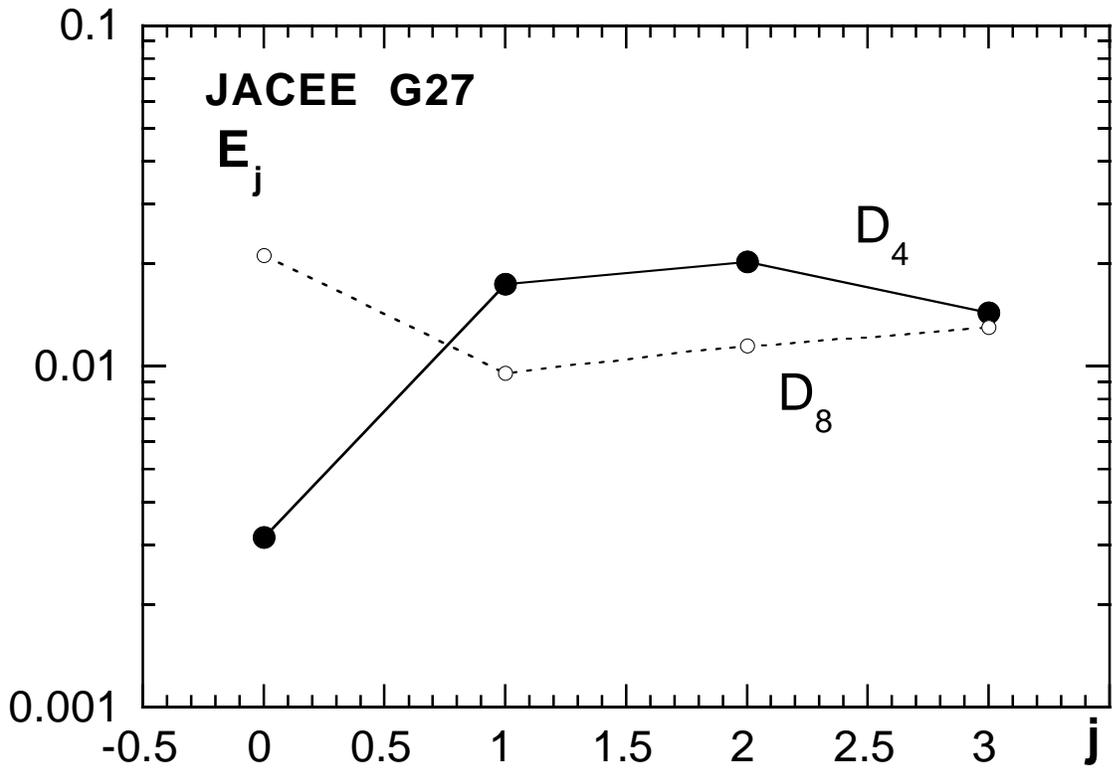
\end{document}